\documentclass[12pt,conference]{IEEEtran}

\usepackage{weiwAlgorithm}
\usepackage{graphicx,subfigure}
\usepackage{enumerate}
\usepackage{multirow}

\graphicspath{{figures/}}
\renewcommand{\baselinestretch}{1.05}
\newcommand{\tabincell}[2]{\begin{tabular}{@{}#1@{}}#2\end{tabular}}

\begin{document}

\title{APT Encrypted Traffic Detection Method based on Two-Parties and Multi-Session for IoT}

\author{\IEEEauthorblockN{Junfeng Xu, Weiguo Lin, Wenqing Fan}
\IEEEauthorblockA{Communication University of China, Beijing, China, 100085 \\
Email: junfeng@cuc.edu.cn}}

\maketitle

\begin{abstract}
APT traffic detection is an important task in network security domain, which is of great significance in the field of enterprise security. Most APT traffic uses encrypted communication protocol as data transmission medium, which greatly increases the difficulty of detection. This paper analyzes the existing problems of current APT encrypted traffic detection methods based on machine learning, and proposes an APT encrypted traffic detection method based on two parties and multi-session. This method only needs to extract a small amount of features, such as session sequence, session time interval, upstream and downstream data size, and convert them into images. Then convolutional neural network method can be used to realize image recognition. Thus, network traffic identification can be realized too. In the preliminary test of five experiments, this method achieves good experimental results, which verifies the effectiveness of the method.
\end{abstract}

\IEEEpeerreviewmaketitle

\section{Introduction}\label{sec:introduction}
Advanced Persistent Threat (APT) refers to the hidden and persistent process of computer network intrusion, usually for commercial or political motives, targeting specific organizations or countries, and maintaining high concealment for a long time \cite{c1}. APT consists of three elements: advanced, long-term and threat. The advanced implies the use of sophisticated malware and technology to exploit vulnerabilities in the system. The long-term implies that an external force will continuously monitor a specific target and obtain data from it. Threat refers to an attack behavior planned by humans. Carrying out APT traffic detection in enterprise network environment is of great significance in network security domain.

Traditional APT traffic detection methods rely heavily on characteristic string detection, which is based on a set of key strings analyzed and obtained from captured APT samples or traffic, such as domain names, URLs, and specific character sequences. As the detection basis of intrusion detection system (IDS) or intrusion prevention system (IPS), the characteristic string works as IoC (Indicators of Compromise) and directly recognizes and matches threats in the traffic. However, in recent years, as more and more APTs tend to use encryption protocols such as TLS for communication, most of the data in application layer has been encrypted during transmission, and this traditional characteristic string detection method are flopping.

Machine learning method can achieve statistical learning recognition effect based on feature set. In recent years, great progress has been made in the field of encrypted traffic identification. In terms of TLS encrypted traffic identification, a certain degree can be achieved with the help of spatio-temporal features, handshake features, certificate features and other non-ciphertext features to achieve some success in the field of malicious traffic detection. However, at present, many research and products use a session / flow (hereinafter referred to as flow) determined by four tuples (source IP, source port, destination IP, destination port, hereinafter referred to as quadruple) as the basic identification unit. This method is difficult to capture the multi-session communication features which is generally seen in APT communication, and the identification effect in many scenes is limited.

In order to solve the above problems, we propose an APT encrypted traffic detection method based on two parties and multi-session. From the multi-session data of two communication parties in a certain period of time, this method extracts multiple recognizable features and transforms them into image data. Then, using the advantages of deep learning method in image recognition domain, we designed convolution neural network method to realize image recognition and indirectly realize flow identification. We used the encrypted traffic of an APT group and the encrypted traffic of normal network application to carry out the experimental test. The experimental results show that the method has achieved good results in accuracy and false positive rate.

	The reminder of this paper is organized as follows: Section \ref{sec:relatedwork} introduces the relevant work and explains the origin of our ideas; Section \ref{sec:method} introduces the overall technical roadmap of the work; Section \ref{sec:evaluation} demonstrates the experimental results and analysis; Finally we conclude in Section \ref{sec:conclusion}.

\section{Related Work}\label{sec:relatedwork}

The traditional APT traffic detection method that relies on feature string matching is relatively mature in the industry and are used in many IDS or IPS products. For example, the Snort \cite{c2} in early stage and Suricata \cite{c3} in recent are based on matching a set of custom rules to achieve their detection function. Such products often rely on the deep packet inspection (DPI), which uses protocol parsing to extract metadata from network flow, and take the metadata as detection unit, Finsterbusch et al. \cite{c4} summarized the current traffic identification methods based on DPI.

In recent years, there have been a lot of research on APT traffic detection method based on machine learning. Anderson et al. published relevant research results in 2016 \cite{c5} and 2017 \cite{c6}. They used various machine learning algorithms such as random forest to carry out threat detection for encrypted traffic. At present, some commercial products based on this method have appeared, such as Cisco's StealWatch \cite{c7} and Huawei's Agile Campus Network \cite{c8}, which can achieve certain practical effects in specific application scenarios. The above research results all take a single session as the identification unit. In the process of APT attack, it contains some features that can only be presented among multiple session, which cannot be seen in a single session. For example, in Command \& Control (C\&C) stage \cite{c9}, there will be many heartbeat sessions or multiple secret stealing sessions between the two communication parties in a certain period of time. These sessions have many identifiable features in terms of interaction sequence, data size, upload/download ratio, etc. However, the current machine learning methods have not made full use of these features.

Based on the above analysis, we propose an APT encrypted traffic detection method based on two parties and mluti-session, trying to make full use of the multi-session features that are not yet used in the current research. Thus, more accurate detection results can be achieved.

\section{Methodology} \label{sec:method}

Since the traffic detection method we proposed has special requirements for training data, we will first introduce the specially created data set, then explain the traffic image conversion method, and finally introduce the CNN model architecture we used.

\subsection{Data Set}

The detection method in this paper needs multiple session data between two communication parties of the same application type. In the public traffic analysis data set, most of them are sorted characteristic data, such as the classic KDD CUP1999 \cite{c10}. In a few datasets that provide raw traffic data, such as USTC-TFC-2016 \cite{c11}, after analysis and comparison, we have not found a data set that can meet the requirements of our method. In order to carry out the experimental work, we used internally collected data to construct traffic data sets that meet the conditions. The data sets consists of two parts: one is the traffic data of an APT group, and specific in command and control phase which generally has more sessions; the second part is four types of normal application traffic data, including browser, mail, Microsoft Office and video. The traffic sessions of each type are grouped according to the both communication parties. The APT traffic set is 3500 groups, and the normal traffic set is 5000 groups each. The details are shown in Table 1.

\begin{table}
    \centering
    \caption{Content of Traffic Data Set}
        \begin{tabular}{c c c c}
          \hline
          Label & Type & \tabincell{c}{Stage \\Application } & \tabincell{c}{Number of \\Session Groups} \\
          \hline
          APT Flow & APT Group &  C\&C & 3500 \\
          \multirow{4}*{Normal Flow} & Browser & Chrome & 5000 \\
          ~ & Mail & Outlook & 5000 \\
          ~ & Office & Excel & 5000 \\
          ~ & Video & Youku & 5000 \\
          \hline
        \end{tabular}
\end{table}

\subsection{Traffic Image Conversion}

Converting the original traffic data to image data which needs to be classified, four steps are needed: traffic analysis, session grouping, feature extraction and image conversion.

\subsubsection{Traffic Analysis}
Traffic analysis is the basic work of traffic classification. Traffic is a kind of continuous data, which needs to be divided into discrete data according to certain rules, and then classification work can be processed on it. At present, the mainstream method is to divide the traffic into multiple session data according to the four tuples, and treat each session as an independent data unit for classification. According to our technical roadmap, we also need to divide the traffic into multiple sessions first. After this step, the input continuous traffic data can be converted into a set of discrete data units composed of multiple sessions. Suppose the input traffic is $T$, then the output data is the session set $S={s_1, s_2, s_3, ..., s_n}$, where s1 to $s_n$ are the data of each session and n is the total number of sessions.

\subsubsection{Session Grouping}
The set of session units obtained from the traffic analysis step are further grouped according to both sides of the communication, where the communication parties are grouped into the triplet of the IP addresses of both parties and the server port. Compared with the four tuples of traffic analysis, the only difference is that the client port information is ignored. A typical scenario is that when a user uses a browser to continuously visit the same HTTPS website within a fixed period of time, multiple TLS protocol sessions will be generated. The client ports of these sessions are different, due to they are randomly selected each time. However, since the web server has a constant IP address, the client accessing the website has also a constant IP address, and the server port is fixed port $443$. Therefore, these sessions can be grouped into one set. Intuitively, these sessions should have similar data properties. Likewise, there are similar scenarios in APT encrypted traffic. For example, the heartbeat sessions used to inform the command and control server that it is alive has similar properties. After this step, the input is session set $S={s_1, s_2, s_3, ..., s_n}$, the output is $G={g_1, g_2, ..., g_m}$, where g1 to gm are session groups, each group contains several sessions. Sessions are arranged in chronological order of their first frame, and m is the number of session groups, or the number of two parties.

\subsubsection{Feature Extraction}
Each session group in the session grouping result is the basic data unit for the subsequent traffic classification and identification. The feature extraction step extracts a set of features for each unit, that will be used in subsequent image conversion steps. The features extracted include session temporal relation, session time interval and up/down data ratio. Session temporal relation refers to the sequence of sessions. The extraction process of session temporal relation is relatively simple. The first frame data of all sessions in each group can be used directly. Thus, the output result is a set of time series data. Session interval refers to the time interval between each two sessions in a group, specifically the time interval between the last frame timestamp of one session and the first frame timestamp of the following session. The up/down data ratio refers to the ratio of the bytes sent from client to server and conversely bytes sent from server to client. Since in the TLS protocol, it is the application layer data that truly reflects the data exchange process, therefore, only the application layer data is used for calculation, that is, the data actually transmitted by the two parties after the key negotiation of the TLS protocol. After this step, the input is session group data $G={g_1, g_2, ..., g_m}$, output as feature set data $F={f_1, f_2, ..., f_m}$, where $f_i$ is feature set, and each set contains three types of feature data, such as session temporal relation, session time interval and up/down data ratio.

\subsubsection{Image Conversion}
Image conversion refers to converting the output from feature extraction step into images, and visually reflecting the three types of feature data mentioned above. An example of the image is shown in Figure 1.

\begin{figure}
  \centering
  \includegraphics[width=3.4in]{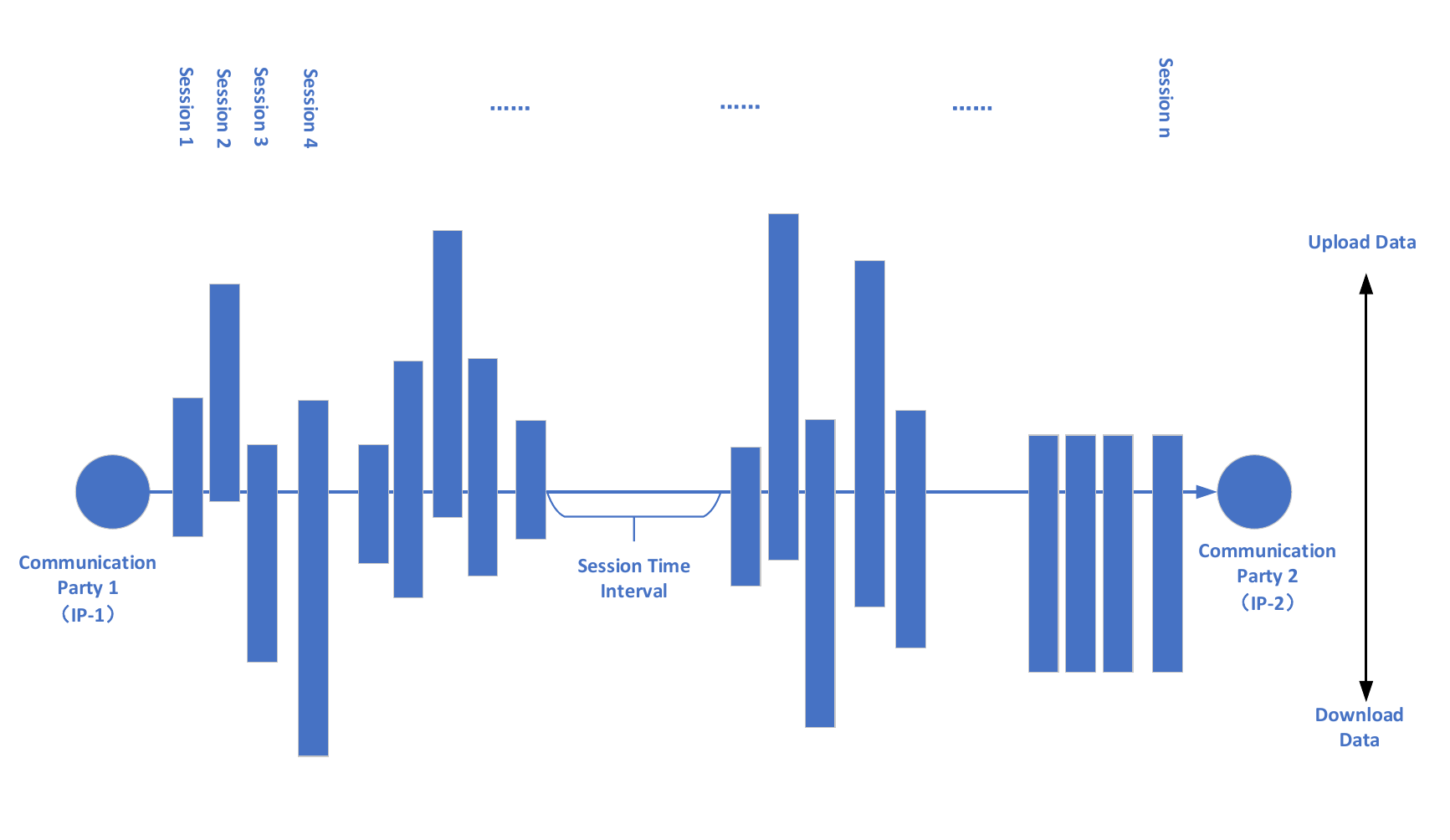}
  \caption{Example of Two-parties Multi-session Image}\label{fig:processstatus}
\end{figure}

The above figure shows the multi-session traffic image of two communication parties. Two ends represent the two communication parties. The left side is the client, using client IP to label, while the right side is the server, using both the server IP and server port to label. The column chart in the middle is the interactive data between the two parties, and each column represents a session. The upper and lower parts of each column on the horizontal axis are up and down data. The length above the horizontal axis represents the bytes of application layer data sent from the client to the server, and the length below represents the bytes of application layer data sent from the server to the client. The order of columns is the order of sessions and the interval between the columns reflect the session time interval. In summary, each pair of communication parties produce one of the above images. Intuitively analyzed, different types of traffic can reflect certain image features for distinguishing and identifying.

\subsection{Convolution Neural Network Architecture}
CNN(Convolutional Neural Network) is currently the mainstream deep learning model in the field of image classification, and has achieved excellent results in many application scenarios. Considering the complexity of the image itself and the amount of training data samples, we use the convolutional neural network architecture of LeNet-5 \cite{c12}, and the architecture is shown in Figure 2.

\begin{figure}
  \centering
  \includegraphics[width=3.4in]{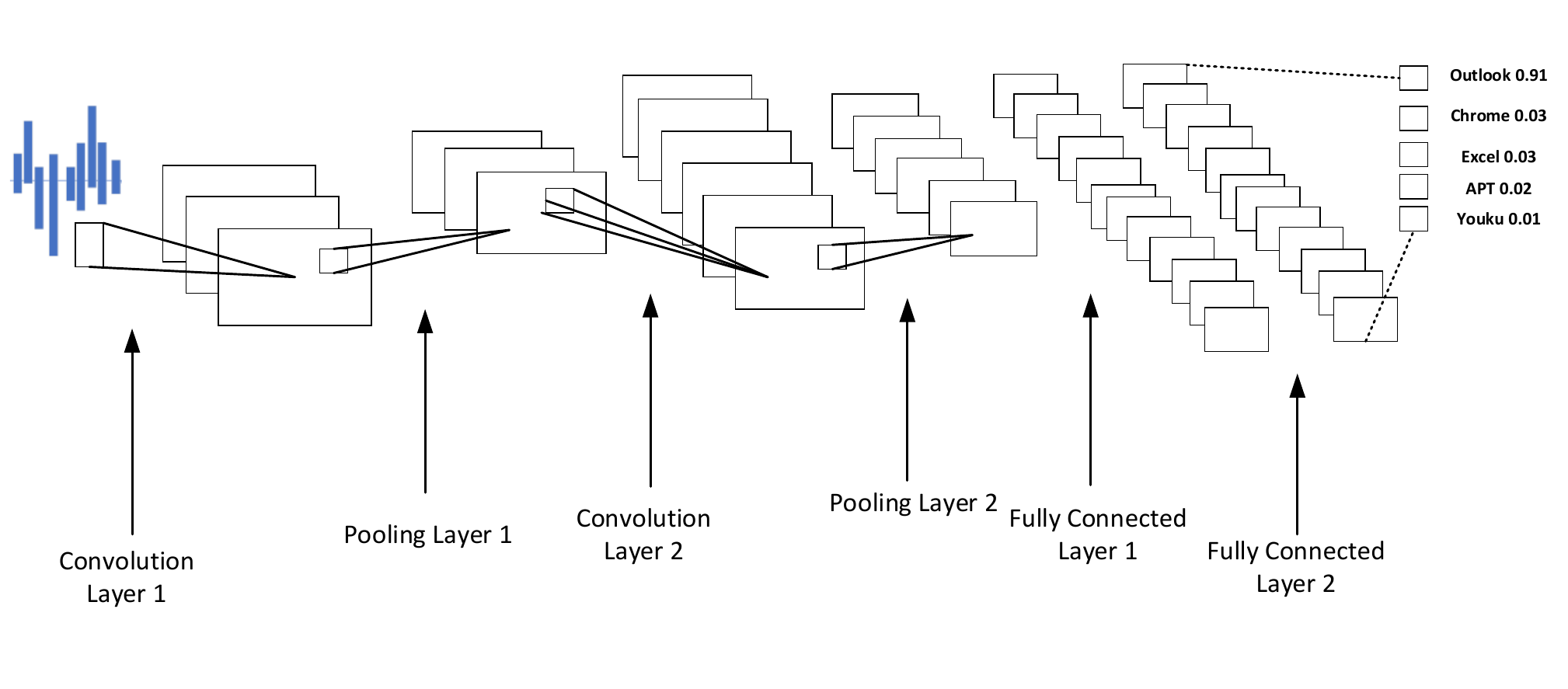}
  \caption{CNN Model Architecture}\label{fig:CNN}
\end{figure}

CNN reads the pixel values from the image file, and these pixel values are normalized and converted from $0~255$ to $0~1$. In the first convolution layer $C_1$, the input is convolved by a convolution kernel with a size of $5*5$. There are $32$ channels to generate $32$ feature maps with size of $28*28$, and then $a 2*2$ maximum pooling operation is performed in $P_1$ layer to generate $32$ feature maps with size of $14*14$. In the second convolution layer $C_2$, the convolution kernel size is also $5*5$, but there are $64$ channels to generate $64$ feature maps with size of $14*14$. Then, $a$ $2*2$ maximum pool operation is performed in $P_2$ layer to generate $64$ $7*7$ feature maps. Next are two fully connected layers, which convert the data size to $1024$ and $10$ in turn. Finally, a $Softmax$ function is used to output various probability values. To reduce overfitting, dropout is used before the output layer.

\section{Evaluation}\label{sec:evaluation}
We use Keras and TensorFlow as the training platform, running on the $Ubuntu 18.04 64-bit$ operating system. $2 / 10$ of the training data was randomly selected as the test data, and the remaining $8 / 10$ was used for training. We used the following criteria to evaluate the proposed method: $Accuracy (A)$, $Precision (P)$, $recall (R)$, and $F_1-Score (F_1)$
as follows:
\begin{equation}\label{A}
  A =\frac{TP + TN}{TP + FP + FN + TN}
\end{equation}

\begin{equation}\label{P}
  P =\frac{TP}{TP + FP}
\end{equation}

\begin{equation}\label{R}
  R =\frac{TP}{TP + FN}
\end{equation}

\begin{equation}\label{F}
  F_1 =\frac{2PR}{P + R}
\end{equation}

Among them, true positive $TP$ represents the number of correctly identified target flows, positive and negative $TN$ denotes the number of other flows correctly identified, false positive $FP$ represents the number of target flows wrongly identified, and false negative $FN$ represents the number of target flows missed.

The experimental results are shown in Table 2. The precision, recall and F1 values refer to the corresponding result data of APT-C\&C flow.

\begin{table*}
    \centering
    \caption{experimental result data (percentage)}
        \begin{tabular}{c c c c c c}
          \hline
          ~ & Data Set & Accuracy & Precision & \tabincell{c}{Recall \\Ratio } & \tabincell{c}{F1 \\Value} \\
          \hline
          \multirow{2} *{1} & APT-C\&C & \multirow{2} *{95.3} & \multirow{2} *{96.3} & \multirow{2} *{92.9} & \multirow{2} *{94.6} \\
          ~ & Chrome & ~ & ~ & ~ & ~ \\

          \multirow{2} *{2} & APT-C\&C & \multirow{2} *{99.8} & \multirow{2} *{99.6} & \multirow{2} *{99.9} & \multirow{2} *{99.7} \\
          ~ & Outlook & ~ & ~ & ~ & ~ \\

          \multirow{2} *{3} & APT-C\&C & \multirow{2} *{92.0} & \multirow{2} *{89.0} & \multirow{2} *{96.5} & \multirow{2} *{92.6} \\
          ~ & Excel & ~ & ~ & ~ & ~ \\

          \multirow{2} *{4} & APT-C\&C & \multirow{2} *{92.1} & \multirow{2} *{97.1} & \multirow{2} *{91.1} & \multirow{2} *{94.0} \\
          ~ & Youku & ~ & ~ & ~ & ~ \\

          \multirow{2} *{5} & APT-C\&C & \multirow{2} *{96.1} & \multirow{2} *{97.0} & \multirow{2} *{95.5} & \multirow{2} *{96.2} \\
          ~ & Mixed Normal Flow & ~ & ~ & ~ & ~ \\

          \hline
        \end{tabular}
\end{table*}

From the result data, we can see that the experiment has achieved good results in the five binary classification tasks, and all the accuracy rate are above 90\%. Among them, the highest 99.8\% results were achieved in the APT-C\&C and Outlook classification experiments. Especially in the fifth experiment, the normal flow set is random mixed data of four normal flow types, which is more closer to the actual application scenario, and the accuracy rate is 96.1\%. In conclusion, through the above experimental results, the effectiveness of our proposed APT encrypted traffic identification method based on two parties and multi-session is preliminarily verified.

\section{Conclusion}\label{sec:conclusion}
In order to solve the problem of APT encrypted traffic identification in the field of network security, an APT encrypted traffic identification method based on two parties and multi-session is proposed. This method does not need complex feature engineering work, but only needs to extract the multi-session temporal relation, time interval and up/down data ratio. Then convert them into image data, realize image recognition by designing convolutional neural network model, and further realize flow identification. The experimental results show that the method has achieved good results in a number of binary classification experiments and verified the effectiveness. In the next stage of our work, we will use more types of data to carry out the verification work, and expand the experimental task to multi-classification scenarios to further explore the application potential of this method.

\section*{Acknowledgment}

This work is supported by the Fundamental Research Funds for the Central Universities(2018XNG1815) and MCM20180504.

\end{document}